\documentclass[a4paper,fleqn]{cas-dc}
\pdfoutput=1
\usepackage{dblfnote}
\usepackage[authoryear,colon,sort&compress]{natbib}

\usepackage{xcolor}
\usepackage{hyperref}
\usepackage{algorithm}
\usepackage{algorithmic}
\usepackage{amsthm}
\usepackage{graphicx}
\hypersetup{
  colorlinks=true,
  linkcolor=blue,
  urlcolor=blue,
  citecolor=blue,
  linktoc=all
}
\def\tsc#1{\csdef{#1}{\textsc{\lowercase{#1}}\xspace}}
\tsc{WGM}
\tsc{QE}
\tsc{EP}
\tsc{PMS}
\tsc{BEC}
\tsc{DE}

\newtheorem{theorem}{Theorem}
\newtheorem{lemma}{Lemma}
\newtheorem{definition}{Definition}
\newtheorem{assumption}{Assumption}
\theoremstyle{remark}
\newdefinition{remark}{Remark}
\newproof{pf}{Proof}
\newproof{pot}{Proof of Theorem \ref{thm}}

\begin{document}
\let\WriteBookmarks\relax
\def\floatpagepagefraction{1}
\def\textpagefraction{.001}
\let\printorcid\relax

\shorttitle{SPTC-AN-FOZNN for solving TVQPs}

\shortauthors{Y Yang et~al.}

\title[mode = title]{A strictly predefined-time convergent and anti-noise fractional-order zeroing neural network for solving time-variant quadratic programming in kinematic robot control}  



\author[1,2]{Yi Yang}

\author[3]{Xiao Li}

\author[1,2]{Xuchen Wang}

\author[1,2]{Mei Liu}

\author[4]{Junwei Yin}

\author[5]{Weibing Li}

\author[6]{Richard M. Voyles}

\author[1,2]{Xin Ma}[type=editor,
   auid=000,bioid=2,
   ]
\cormark[1] 
\ead{xinma001@cuhk.edu.hk} 

\address[1]{Multi-Scale Medical Robotics Center, The Chinese University of Hong Kong, Hong Kong 999077, China}
\address[2]{Department of Mechanical and Automation Engineering, The Chinese University of Hong Kong, Hong Kong 999077, China}
\address[3]{Department of Mechanical and Electrical Engineering, China University of Petroleum (East China), Qingdao 266580, China}
\address[4]{School of Mechanical Engineering, Dalian Jiaotong University, Dalian 116028, China}
\address[5]{School of Computer Science and Engineering, Sun Yat-sen University, Guangzhou 510006, China}
\address[6]{School of Engineering Technology, Purdue University, West Lafayette, IN 47907, USA}

\cortext[1]{Corresponding author} 

\begin{abstract}
This paper proposes a strictly predefined-time convergent and anti-noise fractional-order zeroing neural network (SPTC-AN-FOZNN) model, meticulously designed for addressing time-variant quadratic programming (TVQP) problems. This model marks the first variable-gain ZNN to collectively manifest strictly predefined-time convergence and noise resilience, specifically tailored for kinematic motion control of robots. The SPTC-AN-FOZNN advances traditional ZNNs by incorporating a conformable fractional derivative in accordance with the Leibniz rule, a compliance not commonly achieved by other fractional derivative definitions. It also features a novel activation function designed to ensure favorable convergence independent of the model's order. When compared to five recently published recurrent neural networks (RNNs), the SPTC-AN-FOZNN, configured with $0<\alpha\leq1$, exhibits superior positional accuracy and robustness against additive noises for TVQP applications. Extensive empirical evaluations, including simulations with two types of robotic manipulators and experiments with a Flexiv Rizon robot, have validated the SPTC-AN-FOZNN’s effectiveness in precise tracking and computational efficiency, establishing its utility for robust kinematic control.
\end{abstract}

\maketitle

\section{Introduction}
Quadratic programming (QP) problems are central to diverse scientific and engineering domains, particularly artificial intelligence and robotic kinematic control (\citealp{ref1,ref2,ref3}). These challenges are often formulated as time-variant quadratic programming (TVQP) problems (\citealp{ref1,ref4}). The ubiquity of QPs has catalyzed the development of advanced algorithms for their resolution (\citealp{ref5}). Recurrent neural networks (RNNs), known for their time-series processing capability, have become prominent solvers for these problems due to their ability to operate in parallel hardware realizations (\citealp{ref6,ref7,ref8}). However, traditional Gradient Neural Networks (GNNs), a subset of RNNs, are typically aligned with time-invariant problems and have shown limitations like lagging-behind errors when applied to TVQP challenges (\citealp{ref7,ref9}).

In response to these deficiencies, Zeroing Neural Network (ZNN) models have been developed, providing a reliable framework for solving TVQP problems (\citealp{ref10,ref11}). ZNN models have exhibited exponential convergence in kinematic control of robots, with the choice of activation functions significantly affecting their convergence performance (\citealp{ref12,ref13}). This has led to the establishment of finite-time, fixed-time, and predefined-time convergent ZNNs (\citealp{ref8,ref13,ref14,ref15,ref16,ref17,ref18,ref19}). Moreover, variable gains have been identified as a factor influencing the convergence rate and precision of ZNNs (\citealp{ref20}). Nevertheless, the use of large constant gains or progressively increasing variable gains poses significant challenges in hardware implementations, primarily due to limited power budgets and other practical constraints (\citealp{ref9}). Additionally, the susceptibility of ZNNs to noises (\citealp{ref16,ref21}), such as measurement and computational round-off inaccuracies in hardware realizations, further complicates their deployment in real-world settings, thus highlighting a crucial gap in their practicality and robustness under variable-gain conditions. 

Building upon the distinctive characteristics of the PTC-FOZNN model (\citealp{ref22}), which shows enhanced convergence traits under a time-shrinking gain with a suitably designed activation function, this paper proposes, for the first time, the strictly predefined-time convergent and anti-noise fractional-order zeroing neural network (SPTC-AN-FOZNN) model to address TVQP problems. The SPTC-AN-FOZNN innovatively incorporates a novel activation function that ensures noise resilience and strictly predefined-time convergence, independent of the model’s order. Numerical validations highlight the SPTC-AN-FOZNN’s superior convergence performance compared to five recently established RNN models. Furthermore, its utility as an inverse-kinematics solver for robotic motion planning tasks has been strictly confirmed. The paper’s contributions include:

(1) This research signifies a measured advancement in the development of strictly predefined-time convergent and noise-resilient activation functions for a class of hardware-friendly variable-gain ZNNs, known in our framework as FOZNNs. This results in a novel SPTC-AN-FOZNN model for successful tackling of TVQP problems.

(2) A rigorous proof of the SPTC-AN-FOZNN's strictly predefined-time convergent and noise-tolerant characteristics, demonstrating its practical applicability in kinematic control of robotic manipulators.

(3) Superior convergence precision and robustness against additive noises of the SPTC-AN-FOZNN when compared to five other RNN models, validated through an illustrative numerical example and complementary empirical evaluations, including simulations and experiments with a Franka Emika Panda robot and a Flexiv Rizon robot.

The paper is organized as follows: Section 2 discusses TVQP problems and relevant preliminary theory in the formulation of traditional RNNs. Section 3 details the SPTC-AN-FOZNN and its convergence analysis. Section 4 presents numerical validations, showcasing the model’s efficacy for TVQPs. Section 5 applies the SPTC-AN-FOZNN to robotic kinematic control, illustrating its practical application through simulations and experiments. Section 6 concludes the paper.

\section{Preliminary theory}
\subsection{Problem formulation and preliminaries}
A continuous-time RNN model can be mathematically represented as follows (\citealp{ref23,ref24}):
\begin{equation}
\dot{x}=\mathcal{R}(x(t), t), t \in[0, \infty)
\label{eq1}
\end{equation}
where $x(t)\in\mathbb{R}^n$ denotes the state vector of the system, and $x_0=x(0)$ specifies the system’s initial state. The function $\mathcal{R}(\cdot)$ represents a proper dynamic functional. Assume that the origin, $x(t)=0$ serves as the equilibrium state of the system, the following definitions relevant to the convergence theory can be outlined,
\begin{definition}
(\citealp{ref25}) The origin of the system (\ref{eq1}) is defined as locally finite-time stable if a nonempty open set $\Omega$ around the origin, along with a locally bounded settling-time function $T:\Omega\backslash\{0\}\rightarrow \mathbb{R}_+\cup\{0\}$ exists to ensure that any trajectory $x(t,x_0 )$ originating from an initial state $x_0\in\Omega\backslash\{0\}$ converges to the origin for all $t\ge T(x_0 )$. 
\end{definition}

\begin{definition}
(\citealp{ref26}) The origin of the system (\ref{eq1}) is defined as locally fixed-time stable when it is not only locally finite-time stable but also satisfies the condition where a constant $T_{\mathrm{max}} >0$ exists such that $T(x_0 )\leq T_{\mathrm{max}}$  for all $x_0\in\Omega$. Furthermore, the origin of system \ref{eq1} is said to be globally fixed-time stable if $\Omega=\mathbb{R}^n$. 
\end{definition}

\begin{definition}
(\citealp{ref27}) The origin of the system (\ref{eq1}) with a predefined constant $t_c>0$ is defined as locally predefined-time stable when it is not only locally fixed-time stable but also satisfies the condition where $T(x_0)\leq t_c$ for all $x_0\in\Omega$. Furthermore, the origin of system (\ref{eq1}) is said to be globally predefined-time stable if $\Omega=\mathbb{R}^n$. 
\end{definition}

\begin{definition}
(\citealp{ref14}) The origin of the system (\ref{eq1}) is termed as weakly predefined-time stable if it is predefined-time stable with $T(x_0)\leq t_c$ for all $x_0\in\Omega$. 
\end{definition}

\begin{definition}
(\citealp{ref14}) The origin of the system (\ref{eq1}) is termed as strictly predefined-time stable if it is predefined-time stable with $\sup_{x_0\in\Omega}T(x_0)=t_c$. 
\end{definition}

\begin{remark}
    According to the above definitions, it is known that an RNN model that achieves predefined-time convergence (PTC) also satisfies fixed-time convergence (FIXTC) and finite-time convergence (FNITC) conditions. The strictly predefined-time convergence (SPTC) characteristic of the RNN model guarantees consistent performance and dependability. In constrast to the weakly PTC (WPTC), the SPTC guarantees that the model converges exactly at a user-prescribed time $t_c$.

    \textcolor{blue}{The integration of noise resistance with predefined-time convergence in the RNN model enhances its robustness against disturbances and sensor noise, critical in dynamic settings. This robustness ensures that the RNN consistently meets strict timing constraints and maintains reliable performance, even under variable conditions. Such capabilities are essential in precision-dependent applications like robotic surgery and autonomous vehicle navigation, where safety and effectiveness hinge on accurate timing and stable operation. Together, these features significantly broaden the practical utility and reliability of RNN models.}
\end{remark}

Given the reasons in Remark 1, this work explores a strictly predefined-time convergent and anti-noise recurrent neural solution to the following TVQP problem: 
\begin{equation}
\begin{array}{ll}
\min & x^\mathrm{T}(t) H(t) x(t) / 2+\rho^\mathrm{T}(t) x(t) \\
\text { s.t. } & A(t) x(t)=b(t) \\
& C(t) x(t) \leq d(t)
\end{array}
\label{eq2}
\end{equation}
where $H(t)\in\mathbb{R}^{n\times n}$ is positive semi-definite, $A(t)\in\mathbb{R}^{m\times n}$ and $C(t)\in\mathbb{R}^{p\times n}$ are matrices of full row rank, and $\rho(t)\in\mathbb{R}^n$, $b(t)\in\mathbb{R}^m$ and $d(t)\in\mathbb{R}^p$ are vectors of proper dimensions. If the solution to (\ref{eq2}) exists, it is termed as the Karush-Kuhn-Tucker (KKT) point for the TVQP problem (\ref{eq2}).

\begin{lemma}
    (\citealp{ref28}) $x^\ast(t)\in\mathbb{R}^n$ is the KKT point for the TVQP problem (\ref{eq2}) if for every $\tau\rightarrow0_+$ there exist the Lagrangian multipliers $\phi^\ast(t)\in\mathbb{R}^m$ and $\varphi^\ast(t)\in\mathbb{R}^p$ satisfying
    \begin{equation}
    \left\{\begin{array}{l}
    H(t) x^*(t)+\rho(t)+A^\mathrm{T}(t) \phi^*(t)+C^\mathrm{T}(t) \varphi^*(t)=0 \\
    A(t) x^*(t)-b(t)=0 \\
    \psi_{F B}^\tau\left(d(t)-C(t) x^*(t), \varphi^*(t)\right)=0
    \end{array}\right.
    \label{eq3}
    \end{equation}
    where $\psi_{FB}^{\tau}$ represents the perturbed Fischer-Burmeister (FB) function.
\end{lemma}
The perturbed FB function is defined as (\citealp{ref28}):
\begin{equation}
\psi_{F B}^\tau(u, v)=u+v-\sqrt{u \odot u+v \odot v+\tau}, \tau \rightarrow 0_{+}
\label{eq4}
\end{equation}
where $u,v\in\mathbb{R}^n$ are two vectors of same dimensions. $\tau\in\mathbb{R}^n$ denotes a small perturbation term, the symbol $\odot$ represents the element-wise product operation, and it is worthwhile mentioning the square root also applies to the vector element-wisely.

In accordance with the locally Lipschitz continuity condition detailed in \cite{ref29}, this study asserts the uniqueness of the optimal solution for the TVQP problem (\ref{eq2}). Supported by Lemma 1, the TVQP problem (\ref{eq2}) can be effectively resolved via determining the solution to a time-variant quasi-linear equation (TVQLE):
\begin{equation}
f(y(t),t)=P(t)y(u)+q(t)=0
\label{eq5}
\end{equation}
where $y(t)=[x^\ast{^\mathrm{T}}(t),\phi^\ast{^\mathrm{T}}(t),\varphi^\ast{^\mathrm{T}}(t)]^\mathrm{T}\in\mathbb{R}^{n+m+p}$, and
\begin{equation*}
P(t)=\left[\begin{array}{ccc}
H(t) & A^\mathrm{T}(t) & C^\mathrm{T}(t) \\
A(t) & 0 & 0 \\
-C(t) & 0 & I
\end{array}\right], q(t)=\left[\begin{array}{c}
\rho(t) \\
-b(t) \\
d(t)-n(t)
\end{array}\right]
\end{equation*}
with $n(t)=\sqrt{m(t)\odot m(t)+\varphi^\ast(t)\odot\varphi^\ast(t)+\tau}$ and $m(t)=d(t)-C(t)x^\ast(t)$.

Thus, a vector $y(t)$ satisfying $f(y(t),t)=0$ is recognized as a solution to (\ref{eq3}), and the first $n$ elements of $y(t)$ correspond to the solution for the TVQP problem (\ref{eq2}). The solution to the TVQLE (\ref{eq5}) is achievable through the application of two classes of RNNs, which we will explore in detail in the subsequent section.
\subsection{Traditional GNN and ZNN models}
The following discussion includes an examination of two prevalent RNN models: the GNN model (\citealp{LeCun1998}) and the ZNN model (\citealp{Zhang2002}).
\subsubsection{GNN model}
For the scalar cost function $e(t)=(f(y(t),t))^2/2$, the dynamics of the GNN model is expressed as follows:
\begin{equation}
\dot{y}(t) = -\gamma\nabla e(y)=-\gamma\frac{\partial e(t)}{\partial y(t)}=-\gamma M^\mathrm{T}(t)f(y(t),t)
\label{eq_plus1}
\end{equation}
where $\gamma>0$ denotes the learning rate or gain factor, $\nabla e(y):=\partial e(t)/\partial y(t)$ stands for the gradient of the cost function with respect to $y(t)$, and $M(t)$ represents the coefficient matrix in (\ref{eq7}).

\begin{remark}
  To improve convergence performance, innovative GNN variants have been developed (\citealp{ref15_v1,ref16_v1,ref17_v1}). A prime example is the fractional-order GNN (FO-GNN) model (\citealp{ref17_v1}), which is distinguished by its high accuracy and accelerated convergence rates. Nonetheless, the FO-GNN model presents significant limitations: the convergence time cannot be explicitly predefined by users, and its resistance to additive noise remains inadequate.  
\end{remark}

\begin{remark}
    In scenarios involving time-variant systems, the GNN model (\ref{eq_plus1}) requires frequent recomputation at each time step, which can lead to persistent residual errors or delayed convergence, particularly when the computational iterations are limited (\citealp{ref13}). In contrast, ZNNs employ the evolutionary dynamics involving the derivative information of the vector-valued error function, theoretically eradicates the residual errors common in GNN applications. 
\end{remark}

\subsubsection{ZNN model}
The traditional ZNN model is represented by the RNN described in (\ref{eq6}), which is designed to strategically drive the residual error $\epsilon(t)=f(y(t),t)$ towards zero,
\begin{equation}
\dot{\epsilon}(t) = -\gamma\Phi(\epsilon(t))
\label{eq6}
\end{equation}
 \noindent where $\gamma>0$ represents the learning rate or gain factor, and $\Phi(\cdot):\mathbb{R}^{n+m+p}\rightarrow\mathbb{R}^{n+m+p}$ signifies the activation function.
 
The evolution dynamics in (\ref{eq6}) can be expanded to produce an alternative formulation for the ZNN model, i.e.,
\begin{equation}
M(t)\dot{y}(t) = -N(t)y(t) - \sigma(t) - \gamma\Phi(\epsilon(t))
\label{eq7}
\end{equation}
\noindent where
\begin{equation*}
\begin{aligned}
M(t)&=\left[\begin{array}{ccc}
H(t) & A^\mathrm{T}(t) & C^\mathrm{T}(t) \\
A(t) & 0 & 0 \\
\left(\Pi_1(t)-I\right) C(t) & 0 & I-\Pi_2(t)
\end{array}\right] \\
N(t)&=\left[\begin{array}{ccc}
\dot{H}(t) & \dot{A}^\mathrm{T}(t) & \dot{C}^\mathrm{T}(t) \\
\dot{A}(t) & 0 & 0 \\
\left(\Pi_1(t)-I\right) \dot{C}(t) & 0 & 0
\end{array}\right] \\
\sigma(t)&=\left[\begin{array}{c}
\dot{\rho}(t) \\
-\dot{b}(t) \\
\left(I-\Pi_1(t)\right) \dot{d}(t)
\end{array}\right]
\end{aligned}
\end{equation*}
with $\Pi_1(t) = \mathrm{diag}(m(t)\oslash n(t))$ and $\Pi_2(t)=\mathrm{diag}(\varphi^\ast(t)\oslash n(t))$, where $\oslash$ denotes the Hadamard division and $\mathrm{diag}()$ represents the diagonalization operation.

In hardware realization of an RNN, extraneous noises are inevitably present. A typical noise-perturbed ZNN used for solving the TVQLE (\ref{eq5}) can be characterized as follows,
\begin{equation}
M(t)\dot{y}(t) = -N(t)y(t) - \sigma(t) - \gamma\Phi(\epsilon(t)) + \delta(t)
\label{eq8}
\end{equation}
where $\delta(t)\in\mathbb{R}^{n+m+p}$ represents additive noises, including computational roundoff errors and states' measurement inaccuracies.

In practical scenarios, robotic and mechatronic systems encounter various types of random noise. Typically, computational errors such as truncation and discretization present at lower frequencies, while inaccuracies in state measurement manifest as high-frequency disturbances. Furthermore, the limits of these noises are often not well-defined. Consequently, the following assumption is integral to our analysis:
\begin{assumption}
    The additive noise is presumed bounded, that is, $\|\delta(t)\|_\infty \leq \Delta$, with $\Delta$ representing a constant upper bound for the noise magnitude.
\end{assumption}
\begin{remark}
    This assumption is rooted in the operational norms of conventional mechatronic systems, which are not subject to infinitely large or unbounded noise levels. The noise boundaries are typically specified in machine datasheets, expressed in units like decibels or signal-to-noise ratios, which provide tangible metrics essential for our investigations. \textcolor{blue}{Furthermore, more complex noise conditions, such as mixed-frequency noises, can be treated as linear combinations of the noise conditions discussed above. We have no doubt that the proposed model in our work is also capable of effectively tackling these more complex noisy systems.}
\end{remark}

Prior to exploring the design of the fractional-order neural model, we present a foundational definition and lemma.
\begin{definition}
    (\citealp{ref30}) The conformable fractional derivative of a n-th order differentiable function $h(t)$ is defined as:
    \begin{equation}
    W_\alpha(h)(t) = \lim_{s\rightarrow 0}\frac{h^{(n)}(t+st^{n+1-\alpha})-h^{(n)}(t)}{s}
    \label{eq9}
    \end{equation}
    where $n<\alpha\leq n+1$ represents the order of the conformable fractional derivative operator with $n\ge0$ being an integer.
\end{definition}
A function $h$ is considered $\alpha$-differentiable at a point $t>0$ if the operator $W_\alpha(h)(t)$ is defined at that point. Furthermore, when $0<\alpha\leq1$, the conformable fractional derivative exhibits specific characteristics.
\begin{lemma}
    (\citealp{ref30}) For $0<\alpha\leq1$, if $f$ and $g$ are two $\alpha$-differentiable functions at a point $t>0$, then
    \begin{itemize}
        \item[(a)] $W_\alpha (af+bg)=aW_\alpha (f)+bW_\alpha (g)$, for all $a,b\in\mathbb{R}$.
        \item[(b)] $W_\alpha (t^c )=ct^{c-\alpha}$ for all $c\in\mathbb{R}$.
        \item[(c)] $W_\alpha (C)=0$ for all constant function $f(t)\equiv C$.
        \item[(d)] $W_\alpha (fg)=fW_\alpha (g)+gW_\alpha (f)$.
        \item[(e)] $W_\alpha (f/g)=(gW_\alpha (f)-fW_\alpha (g))/g^2$.
        \item[(f)] $W_\alpha (f)(t)=t^{1-\alpha}f'(t)$ for any $f(t)\in \mathrm{C}^1(-\infty,+\infty)$.
    \end{itemize}
\end{lemma}
\begin{remark}
    Traditional definitions of fractional derivatives, such as Riemann-Liouville and Caputo’s definitions, typically do not adhere to the Leibniz rule (\citealp{10672518,10672519}), impacting the applicability of properties (d), (e), and (f) outlined in Lemma 2. In contrast, the conformable fractional derivative, as defined in (\ref{eq9}), maintains compliance with the Leibniz rule. This adherence enhances its suitability for extending traditional recurrent neural network models into the fractional-order domain. \textcolor{blue}{Moreover, conformable fractional derivatives, unlike traditional definitions that involve complex integral operations, offer a more straightforward formulation that directly generalizes the integer-order derivative (\citealp{10672520}). This simplifies the model’s computational requirements and increases precision by reducing computational complexity and improving accuracy when evaluating transient states between consecutive integer-order states. This efficiency is particularly beneficial in neural network controllers tasked with managing highly dynamic and non-linear system behaviors.}
\end{remark}

\begin{table*}[width=2.1\linewidth,cols=6,pos=t]
\caption{Representative RNN models published recently and their convergence characteristics.}\label{tbl1}
\begin{tabular*}{\tblwidth}{@{} LLLLLL@{} }
\toprule
Model & Gain & Formulation or activation function & Convergence ability & Noise \\
\midrule
\everymath{\scriptstyle}
\multirow{2.5}{*}{FO-GNN} & \multirow{2.5}{*}{$\gamma(t)=\gamma$} & \multirow{2.5}{*}{$\dot{y}(t)=-\gamma \dfrac{M^\mathrm{T}\left(t\right) \epsilon\left(t\right)}{\Gamma(2-\alpha)} \odot\left|y_i-y_{i-1}+\varepsilon\right|^{1-\alpha}$} & \multirow{2.5}{*}{INFTC} & \multirow{2.5}{*}{No} \\

\multirow{2.5}{*}{(\citealp{ref17_v1})} &\multirow{2.5}{*}{} & \multirow{2.5}{*}{} & \multirow{2.5}{*}{} & \multirow{2.5}{*}{} \\
\addlinespace[12pt]


\multirow{2.5}{*}{PRAGNN} & \multirow{2.5}{*}{$\gamma(t)=\gamma$} & \multirow{2.5}{*}{$\dot{y}(t)=-k(t) M^{\mathrm{T}}(t) \epsilon(t)-\gamma_2 M^{\mathrm{T}}(t) \epsilon(t)-\gamma_3 \dfrac{M^{\mathrm{T}}(t) \epsilon(t)}{\left\|M^{\mathrm{T}}(t) \epsilon(t)\right\|}$} & \multirow{2.5}{*}{WPTC} & \multirow{2.5}{*}{Yes} \\

\multirow{2.5}{*}{(\citealp{ref16_v1})} &\multirow{2.5}{*}{} & \multirow{2.5}{*}{} & \multirow{2.5}{*}{} & \multirow{2.5}{*}{}\\
\addlinespace[12pt]


\multirow{2.5}{*}{SPTC-NT-ZNN} & \multirow{2.5}{*}{$\gamma(t)=\gamma$} & \multirow{2.5}{*}{$\Phi(x)=\left\{\begin{array}{l}{x}/{(t_c-t)}, t<t_c \\ x+|x|^p \operatorname{sign}(x)+\xi \operatorname{sign}(x), t \geq t_c\end{array}\right.$} & \multirow{2.5}{*}{SPTC} & \multirow{2.5}{*}{Yes} \\

\multirow{2.5}{*}{(\citealp{ref21})} & \multirow{2.5}{*}{} & \multirow{2.5}{*}{} & \multirow{2.5}{*}{}& \multirow{2.5}{*}{}\\
\addlinespace[12pt]

\multirow{2.5}{*}{NIFZNN} & \multirow{2.5}{*}{$\gamma(t)=\gamma$} & \multirow{2.5}{*}{$\dot{{y}}=-\dfrac{M^\mathrm{T}(t) {\epsilon(t)}}{\left\|M^\mathrm{T}(t) {\epsilon(t)}\right\|^2}\left(\epsilon^{\mathrm{T}}(t) \dfrac{\partial {\epsilon(t)}}{\partial t}+\gamma \dfrac{\|{\epsilon(t)}\|^2}{2}\right)$}& \multirow{2.5}{*}{INFTC} & \multirow{2.5}{*}{No} \\

\multirow{2.5}{*}{(\citealp{Chen2024})} & \multirow{2.5}{*}{} & \multirow{2.5}{*}{} & \multirow{2.5}{*}{}& \multirow{2.5}{*}{} \\
\addlinespace[12pt]

\multirow{2.5}{*}{PTC-FOZNN} & \multirow{2.5}{*}{$\gamma(t)=\gamma t^{\alpha-1}$} & \multirow{2.5}{*}{$\Phi(x)=\dfrac{\pi}{2 \kappa \gamma t_c^\alpha}\left(\|x\|^{1-\kappa}+\|x\|^{1+\kappa}\right) \dfrac{x}{\|x\|}$} & \multirow{2.5}{*}{WPTC} & \multirow{2.5}{*}{No} \\

\multirow{2.5}{*}{(\citealp{ref22})} & \multirow{2.5}{*}{} & \multirow{2.5}{*}{} & \multirow{2.5}{*}{} & \multirow{2.5}{*}{} \\
\addlinespace[12pt]

\multirow{6}{*}{SPTC-AN-FOZNN} & \multirow{6}{*}{$\gamma(t)=\gamma t^{\alpha-1}$} & \multirow{6}{*}{$\Phi(x)=\left\{\begin{array}{l}\dfrac{1}{\gamma t_c^{\alpha-1}}\left(\dfrac{x}{t_c-t}+\dfrac{\gamma\|x\|^2}{\left(t_c-t\right)^2} \dfrac{x}{\|x\|}\right), t<t_c \\ \dfrac{1}{\gamma t^{\alpha-1}}\left(\dfrac{x}{t_p-t}+\left(\zeta+\dfrac{\gamma t^{\alpha-1}\|x\|^2}{\left(t_p-t\right)^2}\right) \dfrac{x}{\|x\|}\right), t \geq t_c\end{array}\right.$}& \multirow{6}{*}{SPTC for $0<\alpha\leq1$} & \multirow{6}{*}{Yes} \\

\addlinespace[55pt]
\bottomrule
\multicolumn{6}{p{0.88\textwidth}}{\footnotesize Note: "INFTC", "SPTC" and "WPTC" means convergence in infinite time, strictly, and weakly predefined time, respectively. Besides, the listed activation function should be combined with (\ref{eq8}) or (\ref{eq12}) to give the formulation of the specific model.}
\end{tabular*}
\end{table*}

\section{Scheme design and convergence analysis}
This section elaborates on the design of the SPTC-AN-FOZNN model. It also provides theoretical justifications to support model’s strictly predefined-time convergence and its noise resilience.
\subsection{Design of SPTC-AN-FOZNN model}
A fractional-order ZNN (FOZNN) model is formulated by substituting $\dot{\epsilon}(t)$ in (\ref{eq6}) with conformable fractional derivative of $\epsilon(t)$, i.e.,
\begin{equation}
W_\alpha(\epsilon)(t)=t^{1-\alpha}\dot{\epsilon}(t)=-\gamma\Phi(\epsilon(t))
\label{eq10}
\end{equation}
with its expanded expression given as
\begin{equation}
M(t)\dot{y}(t) = -N(t)y(t) - \sigma(t) - \gamma t^{\alpha-1}\Phi(\epsilon(t))
\label{eq11}
\end{equation}
where the order $\alpha$ is prescribed within $0<\alpha\leq1$, and the term $t^{\alpha-1}$ is derived from the property (f) associated with the conformable fractional derivative as outlined in Lemma 2. Then, the noise-perturbed FOZNN model is presented as 
\begin{equation}
M(t)\dot{y}(t) = -N(t)y(t) - \sigma(t) - \gamma t^{\alpha-1}\Phi(\epsilon(t)) + \delta(t)
\label{eq12}
\end{equation}

To confer the strictly predefined-time convergence attribute upon the FOZNN model (\ref{eq10}) with or without external noises, we propose the following predefined-time stabilizer to serve as the activation function,
\begin{equation}
\Phi(x)=\left\{\begin{array}{l}
\frac{1}{\gamma t_c^{\alpha-1}}\left(\frac{x}{t_c-t}+\frac{\gamma\|x\|^2}{\left(t_c-t\right)^2} \frac{x}{\|x\|}\right), t<t_c \\
\frac{1}{\gamma t^{\alpha-1}}\left(\frac{x}{t_p-t}+\left(\zeta+\frac{\gamma t^{\alpha-1}\|x\|^2}{\left(t_p-t\right)^2}\right) \frac{x}{\|x\|}\right), t \geq t_c
\end{array}\right.
\label{eq13}
\end{equation}
\noindent where $t_p=(1-\exp(-\pi/(2\sqrt{\zeta(\zeta-\Delta)})))t_c$ with $t_c>0$ being the predefined time, and $\zeta$ being an arbitrary positive number greater than $\Delta$. Then, (\ref{eq11}) and (\ref{eq12}), combined with (\ref{eq13}), establish the SPTC-AN-FOZNN model for solving the TVQP problem (\ref{eq2}).

\subsection{Main results in theoretical analysis}
Theorems presented herein guarantee the strictly predefined-time convergence of FOZNNs (\ref{eq11}) and (\ref{eq12}) under the activation function (\ref{eq13}).

\begin{theorem}
    For the TVQP problem (\ref{eq2}) or equivalently, the TVQLE (\ref{eq5}), if the predefined-time stabilizer (\ref{eq13}) is utilized, the neural state vector $y(t)$ for the non-noise FOZNN model (\ref{eq11}), originating from an initial condition $y_0$ sufficiently close to the theoretical initial state $y_0^\ast$, can converge to the theoretical solution $y^\ast(t)$ in strictly predefined time $t_c$.
\end{theorem}
\noindent\textbf{Proof.} For $t<t_c$, integrating the designed predefined-time stabilizer (\ref{eq13}) into the FOZNN model (\ref{eq11}) yields:
\begin{equation}
\dot{\epsilon}(t)=-\frac{t^{\alpha-1}}{t_c^{\alpha-1}}\left(\frac{\epsilon(t)}{t_c-t}+\gamma \frac{\|\epsilon(t)\|^2}{\left(t_c-t\right)^2} \frac{\epsilon(t)}{\|\epsilon(t)\|}\right)
\label{eq14}
\end{equation}

Design a Lyapunov functional candidate as $V(t)=\|\epsilon(t)\|/(t_c-t)$. For $t<t_c$, the time derivative of $V(t)$ is
\begin{equation}
\begin{aligned}
\dot{V}(t) & =\frac{\epsilon^\mathrm{T}(t)}{\|\epsilon(t)\|\left(t_c-t\right)} \dot{\epsilon}(t)+\frac{\|\epsilon(t)\|}{\left(t_c-t\right)^2} \\
& =-\frac{t^{\alpha-1}}{\left(t_c-t\right) t_c^{\alpha-1}}\left(\frac{\| \epsilon(t)\|}{t_c-t}+\gamma \frac{\|\epsilon(t)\|^2}{\left(t_c-t\right)^2}\right)+\frac{\|\epsilon(t)\|}{\left(t_c-t\right)^2} \\
& \leq-\frac{\gamma V^2}{t_c-t} \leq 0
\end{aligned}
\label{eq15}
\end{equation}

The Lyapunov stability theorem suggests that the origin is globally finite-time stable. Subsequently, integrating on both sides of the differential inequality (\ref{eq15}) yields: 
\begin{equation}
\ln \frac{t_c}{t_c-t}=\int_0^t \frac{\mathrm{d} \tau}{t_c-\tau} \leq \int_{V(0)}^{V(t)} \frac{\mathrm{d} V}{-\gamma V^2}=\frac{1}{\gamma}\left(\frac{1}{V(t)}-\frac{1}{V(0)}\right)
\label{eq16}
\end{equation}

Then, (\ref{eq16}) implies that $V(t)\rightarrow0$ as $t\rightarrow t_c$. Thus, one can obtain that $\|\epsilon(t)\|=(t_c-t)V(t)\rightarrow0$ as $t\rightarrow t_c$.
For $t\ge t_c$, substituting the activation function (\ref{eq13}) into the non-noise FOZNN model (\ref{eq11}) produces:
\begin{equation}
\dot{\epsilon}(t)=-\left(\frac{\epsilon(t)}{t_p-t}+\left(\zeta+\gamma t^{\alpha-1} \frac{\|\epsilon(t)\|^2}{\left(t_p-t\right)^2}\right) \frac{\epsilon(t)}{\|\epsilon(t)\|}\right)
\label{eq17}
\end{equation}

For $t\ge t_c$, we consider a Lyapunov functional candidate $V(t)=\|\epsilon(t)\|/(t-t_p)$,  and obtains that 
\begin{equation}
\begin{aligned}
\dot{V}(t) & =\frac{\epsilon^\mathrm{T}(t)}{\|\epsilon(t)\|\left(t-t_p\right)} \dot{\epsilon}(t)-\frac{\|\epsilon(t)\|}{\left(t-t_p\right)^2} \\
& =\frac{1}{t_p-t}\left(\frac{\|\epsilon(t)\|}{t_p-t}+\gamma t^{\alpha-1} \frac{\|\epsilon(t)\|^2}{\left(t_p-t\right)^2}+\zeta\right)-\frac{\|\epsilon(t)\|}{\left(t-t_p\right)^2} \\
& \leq-\frac{\gamma t^{\alpha-1} V^2}{t-t_p}-\frac{\zeta}{t-t_p}<0
\end{aligned}
\label{eq18}
\end{equation}
which is always negative, indicating that $\epsilon(t)=0$ is strictly maintained for all $t>t_c$. This suggests that the non-noise FOZNN model (\ref{eq11}) with the activation function (\ref{eq13}) exhibits strictly predefined-time convergence. $\;\;\;\;\blacksquare$

\begin{theorem}
    For the TVQP problem (\ref{eq2}) or equivalently, the TVQLE (\ref{eq5}), if the predefined-time stabilizer (\ref{eq13}) is utilized, the neural state vector $y(t)$ for the noise-perturbed FOZNN model (\ref{eq12}), originating from an initial condition $y_0$ sufficiently close to the theoretical initial state $y_0^\ast$, can converge to the theoretical solution $y^\ast(t)$ in strictly predefined time $t_c$.
\end{theorem}
\noindent\textbf{Proof.} For $t<t_c$, substituting the designed predefined-time stabilizer (\ref{eq13}) into the FOZNN model (\ref{eq12}) yields:
\begin{equation}
\dot{\epsilon}(t)=-\frac{t^{\alpha-1}}{t_c^{\alpha-1}}\left(\frac{\epsilon(t)}{t_c-t}+\gamma \frac{\|\epsilon(t)\|^2}{\left(t_c-t\right)^2} \frac{\epsilon(t)}{\|\epsilon(t)\|}\right)+\delta(t)
\label{eq19}
\end{equation}

Design a Lyapunov functional candidate as $V(t)=\|\epsilon(t)\|/(t_c-t)$. For $t<t_c$, the time derivative of $V(t)$ is
\begin{equation}
\begin{aligned}
\dot{V}(t)= & \frac{\epsilon^\mathrm{T}(t)}{\|\epsilon(t)\|\left(t_c-t\right)} \dot{\epsilon}(t)+\frac{\|\epsilon(t)\|}{\left(t_c-t\right)^2} \\
= & -\frac{t^{\alpha-1}}{\left(t_c-t\right) t_c^{\alpha-1}}\left(\frac{\|\epsilon(t)\|}{t_c-t}+\gamma \frac{\|\epsilon(t)\|^2}{\left(t_c-t\right)^2}\right) \\
& \quad+\frac{\epsilon^\mathrm{T}(t) \delta(t)}{\|\epsilon(t)\|\left(t_c-t\right)}+\frac{\|\epsilon(t)\|}{\left(t_c-t\right)^2} \\
& \leq-\frac{\gamma V^2}{t_c-t}+\frac{\Delta}{t_c-t}
\end{aligned}
\label{eq20}
\end{equation}

The Lyapunov stability theorem suggests that the origin is globally finite-time stable. Subsequently, integrating on both sides of the differential inequality (\ref{eq20}) gives: 
\begin{equation}
\begin{aligned}
\ln \frac{t_c}{t_c-t}=\int_0^t \frac{\mathrm{d} \tau}{t_c-\tau} & \leq \int_{V(0)}^{V(t)} \frac{\mathrm{d} V}{\Delta-\gamma V^2} \\
& =\frac{1}{2 \sqrt{\gamma \Delta}} \ln \frac{|(u(0)-1)(u(t)+1)|}{|(u(0)+1)(u(t)-1)|}
\end{aligned}
\label{eq21}
\end{equation}
where $u(t)=V(t)\sqrt{\gamma/\Delta}$. Then, from (\ref{eq21}), one can observe that $u(t)\rightarrow1$ and $V(t)\rightarrow\sqrt{\Delta/\gamma}$ as $t→t_c$. Thus, one can deduce that $\|\epsilon(t)\|=(t_c-t)V(t)\rightarrow0$ as $t→t_c$.

For $t\ge t_c$, substituting the activation function (\ref{eq13}) into the noise-perturbed FOZNN model (\ref{eq12}) produces:
\begin{equation}
\dot{\epsilon}(t)=-\left(\frac{\epsilon(t)}{t_p-t}+\left(\zeta+\gamma t^{\alpha-1} \frac{\|\epsilon(t)\|^2}{\left(t_p-t\right)^2}\right) \frac{\epsilon(t)}{\|\epsilon(t)\|}\right)+\delta(t)
\label{eq22}
\end{equation}

For $t\ge t_c$, we consider a Lyapunov functional candidate $V(t)=\|\epsilon(t)\|/(t-t_p)$,  and obtains that 
\begin{equation}
\begin{aligned}
\dot{V}(t)= & \frac{\epsilon^\mathrm{T}(t)}{\|\epsilon(t)\|\left(t-t_p\right)} \dot{\epsilon}(t)-\frac{\|\epsilon(t)\|}{\left(t-t_p\right)^2} \\
= & \frac{1}{t_p-t}\left(\frac{\|\epsilon(t)\|}{t_p-t}+\gamma t^{\alpha-1} \frac{\|\epsilon(t)\|^2}{\left(t_p-t\right)^2}+\zeta\right)-\frac{\|\epsilon(t)\|}{\left(t-t_p\right)^2} \\
& \quad+\frac{\epsilon^\mathrm{T}(t) \delta(t)}{\|\epsilon(t)\|\left(t-t_p\right)} \\
\leq & -\frac{\gamma t^{\alpha-1} V^2}{t-t_p}-\frac{\zeta-\Delta}{t-t_p}<0
\end{aligned}
\label{eq23}
\end{equation}
which is always negative, indicating that $\epsilon(t)=0$ is strictly maintained for all $t>t_c$. This suggests that the noise-perturbed FOZNN model (\ref{eq12}) with the activation function (\ref{eq13}) exhibits strictly predefined-time convergence. $\;\;\;\;\blacksquare$
\begin{remark}
    The SPTC-AN-FOZNN is evaluated against a range of recently developed RNN models as detailed in Table 1. While not the pioneer in achieving strictly predefined-time convergence and noise resilience—attributes also seen in the SPTC-NT-ZNN model (\citealp{ref21})—the SPTC-AN-FOZNN introduces two notable innovations. Firstly, it employs a fractional-order approach, utilizing a time-diminishing gain factor, $\gamma(t) = \gamma t^{\alpha-1}$. This approach enhances energy efficiency in hardware implementations, especially as the system approaches the steady state, by reducing power consumption as $t \rightarrow \infty$. Secondly, unlike the SPTC-NT-ZNN, whose piecewise activation function may induce fluctuations in residual errors at the transition point $t_c$, the SPTC-AN-FOZNN uses a structurally coherent piecewise function that ensures a smoother transition in error dynamics at critical time points. These enhancements not only contribute to greater energy efficiency but also ensure more stable convergence behavior, underscoring the SPTC-AN-FOZNN's advancement over existing models. \textcolor{blue}{However, it is worthwhile noting that the computational cost of the SPTC-AN-FOZNN model, similar to the SPTC-NT-ZNN and PTC-FOZNN, lags behind the FO-GNN, PRAGNN, and NIFZNN models, which benefit from simpler operations with lower algorithmic complexity.}
\end{remark}
\textcolor{blue}{\begin{remark}
    The strictly predefined-time convergence characteristics speculated in the Theorem 1 and Theorem 2 can be extended to $\alpha>1$ for the SPTC-AN-FOZNN model. Consider $\alpha>1$, the first derivative of $V(t)$ in (\ref{eq15}) becomes:
\begin{equation}
\begin{aligned}
\dot{V}(t) & =-\frac{t^{\alpha-1}}{\left(t_c-t\right) t_c^{\alpha-1}}\left(\frac{\| \epsilon(t)\|}{t_c-t}+\gamma \frac{\|\epsilon(t)\|^2}{\left(t_c-t\right)^2}\right)+\frac{\|\epsilon(t)\|}{\left(t_c-t\right)^2} \\
& \leq -\frac{t}{\left(t_c-t\right) t_c}\left(\frac{\| \epsilon(t)\|}{t_c-t}+\gamma \frac{\|\epsilon(t)\|^2}{\left(t_c-t\right)^2}\right)+\frac{\|\epsilon(t)\|}{\left(t_c-t\right)^2} \\
& \leq \frac{-\gamma tV^2}{(t_c-t)t_c} + \frac{V}{t_c}
\end{aligned}
\label{eq15_c1}
\end{equation}
\indent Note that the above differential inequality implies that $V(t)\rightarrow 0$ as $t\rightarrow 0$. This is because as $t$ gets very close to $t_c$, the first term $-\gamma tV^2/((t_c-t)t_c)$ might dominate due to the small value in the denominator. A key aspect to consider is how $V(t)$ behaves to prevent the term from diverging. If $V(t)$ remains bounded and doesn’t tend to zero, this term could lead to a scenario where $\dot{V}(t)$ becomes very large negatively (since $V^2$ in the numerator could not offset the rapid decrease in $t_c-t$ in the denominator), pushing $V(t)$ to decrease rapidly. However, if $V(t)$ tends to zero faster than $t_c-t$ tends to zero, the term could remain bounded.
\end{remark}}
\begin{figure*}
	\centering
	\includegraphics[scale=.38]{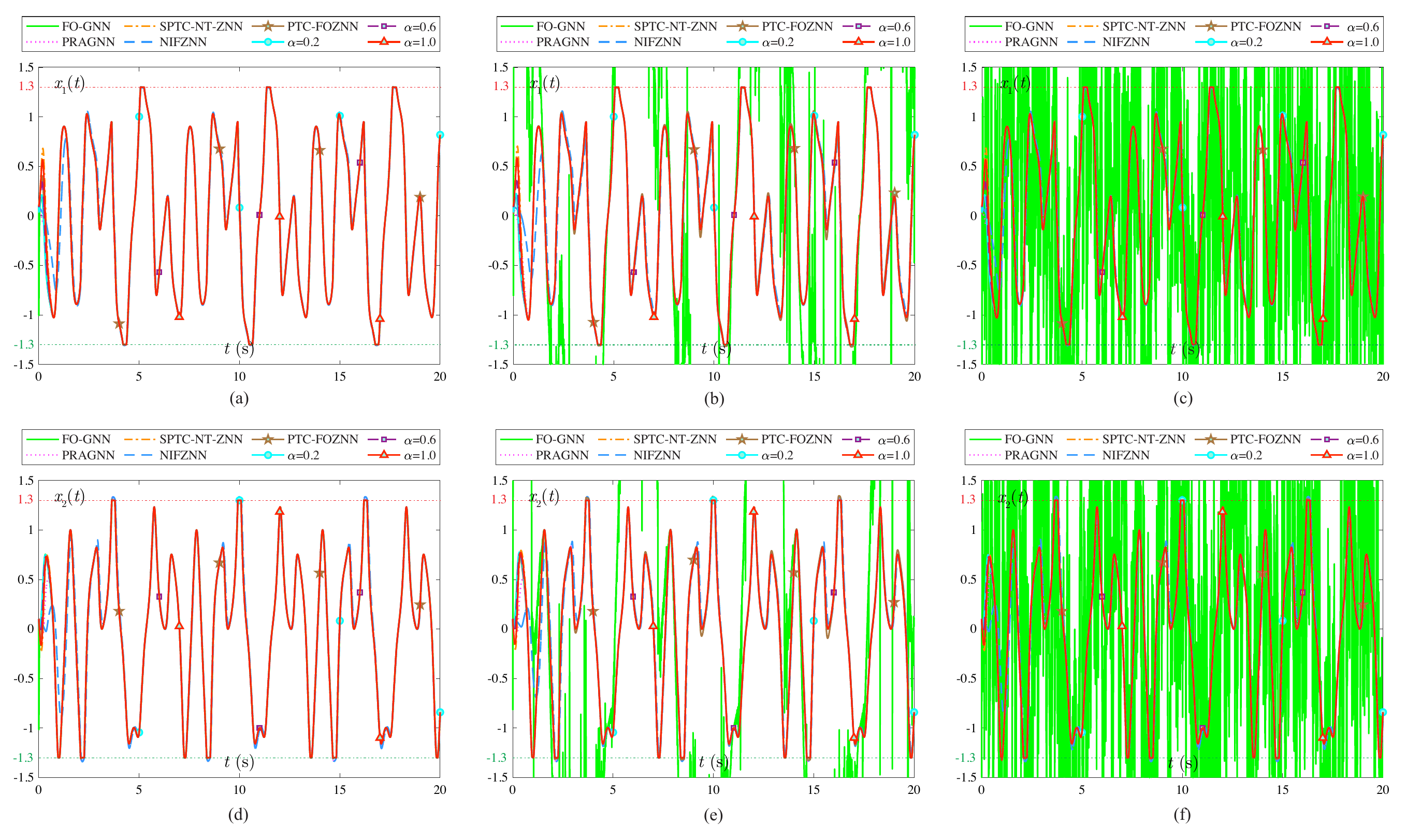}
        \caption{Profiles of neural states $x_1$ and neural state $x_2$ across six models for solving the TVQP problem (\ref{eq25}) under the following noise conditions: (a) and (d) without additive noise; (b) and (e) a low-frequency noise $\delta(t)=0.2\cos(t)$; (c) and (f) a high-frequency random noise $\delta(t)=0.5\bar{n}(t)$, where $\bar{n}(t)$ is a white noise signal bounded by $1$. (Our SPTC-AN-FOZNN model takes three distinct $\alpha$ values)}
	\label{fig1}
\end{figure*}
  
\begin{figure*}
	\centering
		\includegraphics[scale=.38]{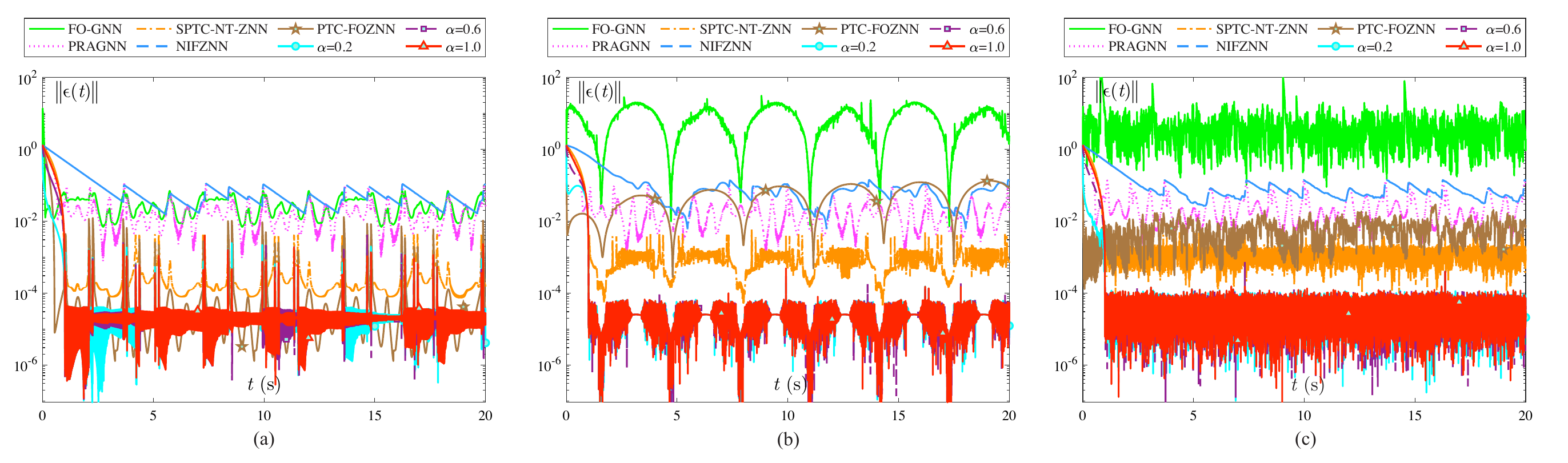}
	\caption{Profiles of the residual error $\|\epsilon(t)\|$ across six models for solving the TVQP problem (\ref{eq25}) under the following noise conditions: (a) without additive noise; (b) a low-frequency noise $\delta(t)=0.2 \cos(t)$; (c) a high-frequency random noism $\delta(t)=0.5\bar{n}(t)$, where $\bar{n}(t)$ is a white noise signal bounded by $1$. (Our SPTC-AN-FOZNN odel takes three distinct $\alpha$ values.)}
	\label{fig2}
\end{figure*}
\section{Numerical validation}
The efficacy of the proposed SPTC-AN-FOZNN model is evaluated through a comparative study with five recent RNN models, including the FO-GNN model (\citealp{ref17_v1}), PRAGNN model (\citealp{ref16_v1}), SPTC-NT-ZNN model (\citealp{ref21}), NIFZNN model (\citealp{Chen2024}), and the PTC-FOZNN model (\citealp{ref22}). This assessment is performed on the following TVQP problem:
\begin{equation}
\begin{aligned}
\min \quad &(\sin (t) / 8+1 / 2) x_1^2(t)+(\cos (t) / 8+1 / 2) x_2^2(t)\\
&+\cos (t) x_1(t) x_2(t) / 2+\cos (3 t) x_1(t)\\
&+\sin (3 t) x_2(t)\\
\text {s.t.} \quad & \cos (4 t) x_1(t)+\sin (4 t) x_2(t)=\sin (2 t)\\
&-1.3 \leq x_1(t), x_2(t) \leq 1.3
\end{aligned}
\label{eq25}
\end{equation}
where the coefficients associated to the compact form (\ref{eq2}) are $A=[\cos(4t),\sin(4t)]$, $b=\sin(2t)$, $d=[1.3,1.3,1.3,1.3]^\mathrm{T}$, $C=[I,-I]^\mathrm{T}$, and
\begin{equation*}
H=\left[\begin{array}{cc}
\sin (t) / 4+1 & \cos (t) / 2 \\
\cos (t) / 2 & \cos (t) / 4+1
\end{array}\right], \rho=\left[\begin{array}{c}
\cos (3 t) \\
\sin (3 t)
\end{array}\right]
\end{equation*}
where $I$ denotes an identity matrix. All six models under consideration are discretized using the forward Euler method, with a discrete step size $\Delta t=1\times10^{-3}$ seconds, ensuring a consistent evaluation framework. The models employ a constant gain factor $\gamma=2$, a predefined time $t_c=1$ s, and parameters $p=0.5$, $\kappa=0.5$, $\Delta=\|\delta(t)\|_\infty$, $\zeta=5\Delta$, $\gamma_2=0$, $\gamma_3=\zeta$, and $\xi=\zeta/\gamma$ are applied to all neural models. Additionally, $\tau=1\times10^{-8}$ is set for the perturbed FB function.

In an enhancement to the standard configuration, our SPTC-AN-FOZNN model introduces a variable gain $\gamma(t)=\gamma t^{\alpha-1}$ where $0<\alpha\leq 1$, aimed at reducing power consumption in hardware implementations through quick mitigation of the model’s gain. In this simulation, we consider the three types of additive noise simulated in the computation and hardware implementation/measurements, namely, $\delta(t)=0$, $\delta(t)=0.2\cos(t)$, and $\delta(t)=0.5\bar{n}(t)$, with $\bar{n}(t)$ being a white noise signal bounded by $1$. The neural state profiles for all models, under the influence of three types of noise conditions, are depicted in Fig. \ref{fig1}. Fig. \ref{fig2} showcases the residual error profiles for the six models, both in noise-free conditions and under the two specified noise scenarios. It is noteworthy that the SPTC-AN-FOZNN model, particularly with three distinct values of $\alpha$, demonstrates superior precision and enhanced noise immunity compared to its counterparts. Notably, this model achieves convergence (e.g., $\|\epsilon(t)\|\sim10^{-5}$) within the strictly predefined time $1$ second for all three noise scenarios, reducing residual error by over $90\%$ relative to the SPTC-NT-ZNN and other RNN models. These findings affirm the SPTC-AN-FOZNN model's exceptional efficacy in addressing general TVQP problems.

\textcolor{blue}{Further analysis reveals that in high dynamic range environments, where signal variability can be intense and unpredictable, the SPTC-AN-FOZNN model maintains optimal performance due to its adaptive gain adjustment, $\gamma(t)$. This adjustment allows the model to respond more efficiently to sudden changes in signal amplitude, thereby ensuring faster convergence rates and substantially reduced error margins. Additionally, in scenarios characterized by rapid system state changes, the model's ability to quickly adapt its gain in response to the state's dynamics proves invaluable. These attributes are particularly beneficial in applications demanding stringent real-time performance and robust noise immunity, such as in autonomous robotics and adaptive signal processing. Such applications often face scenarios where the rapid and accurate adjustment of control actions is critical to system stability and operational success, underscoring the practical relevance and superior performance of the SPTC-AN-FOZNN model.}

\begin{figure}[t]
	\centering
		\includegraphics[scale=.32]{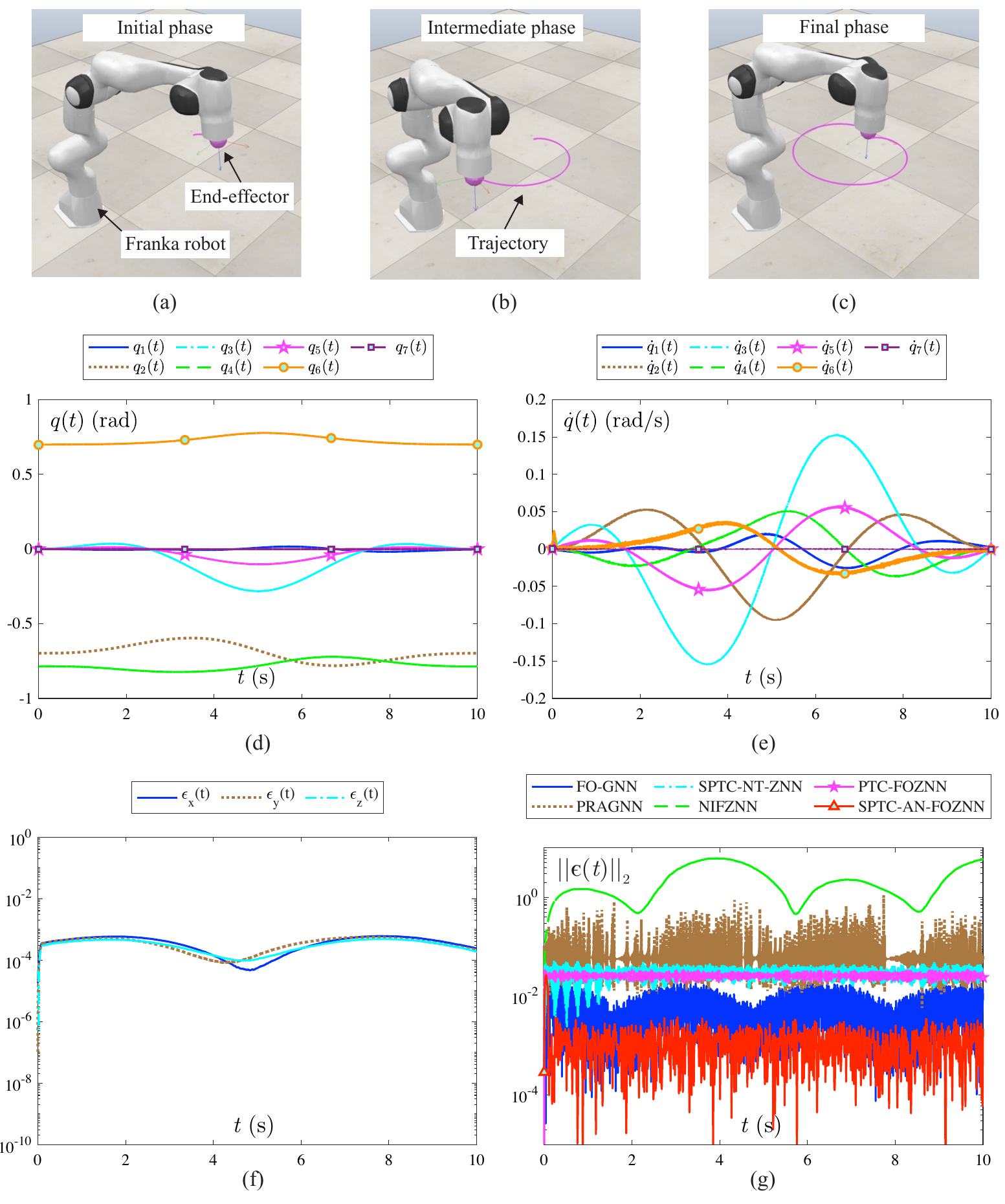}
	\caption{The snapshots of the (a) initial, (b) intermediate, and (c) final phases for a simulated Franka Emika Panda robot during tracking a heart-shaped path, with the rendered (d) joint angles, (e) joint velocities, (f) absolute tracking errors by the SPTC-AN-FOZNN model, and (g) residual errors rendered by six different neural models with a bounded random noise $\delta(t)=\cos(t)+\bar{n}(t)$.}
	\label{fig3}
\end{figure}

\begin{figure}[t]
	\centering
		\includegraphics[scale=.32]{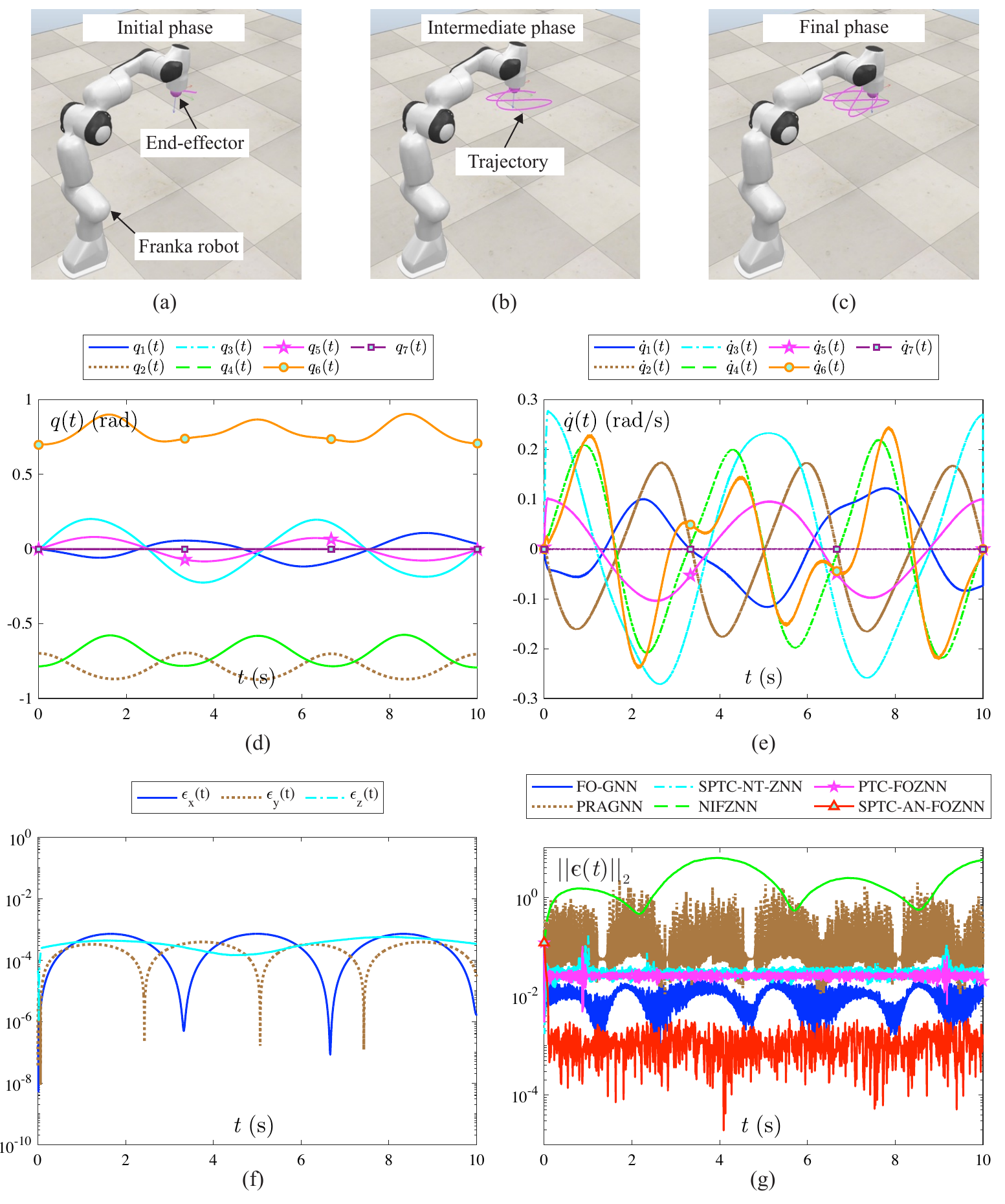}
	\caption{The snapshots of the (a) initial, (b) intermediate, and (c) final phases for a simulated Franka Emika Panda robot during tracking a Lissajous curve, with the rendered (d) joint angles, (e) joint velocities, (f) absolute tracking errors by the SPTC-AN-FOZNN model, and (g) residual errors rendered by six different neural models with a bounded random noise $\delta(t)=\cos(t)+\bar{n}(t)$.}
	\label{fig4}
\end{figure}
\section{Kinematic control of robotic manipulators}
This section presents the application of the SPTC-AN-FOZNN model to the kinematic motion control of robotic manipulators through the formulation of the TVQP problem below (\citealp{ref13,ref31}):
\begin{equation}
\begin{array}{ll}
\min & \dot{q}^\mathrm{T}(t) \dot{q}(t) / 2+\rho^\mathrm{T}(t) \dot{q}(t) \\
\text {s.t.} & J(t) \dot{q}(t)=\dot{w}(t) \\
& q^{-} \leq q(t) \leq q^{+} \\
& \dot{q}^{-} \leq \dot{q}(t) \leq \dot{q}^{+}
\end{array}
\label{eq26}
\end{equation}
where $q(t),\dot{q}(t)\in\mathbb{R}^n$ represent the joint angles and velocities of the robotic manipulator, respectively. The vector $\rho(t)=\iota(q(t)-q_0)$ aids the repetitive motion of the robotic manipulator, with $\iota\in\mathbb{R}$ being a positive constant and $q_0$ the initial joint angle. $J(t)\in\mathbb{R}^{3\times n}$ is the Jacobian matrix, and $w(t)\in\mathbb{R}^3$ signifies the desired trajectory of the end-effector. The bounds $q^-$, $q^+$, $\dot{q}^-$ and $\dot{q}^+$ specify the permissible ranges for joint angles and velocities.

The problem is transformed into the compact form (\ref{eq2}) with $x(t)=\dot{q}(t)$, $H(t)=I$, $A(t)=J(t)$, $b(t)=\dot{w}(t)$, $C(t)=[I,-I]^\mathrm{T}$, and $d(t)=[{d}^{+\mathrm{T}},-{d}^{-\mathrm{T}}]$. According to \cite{ref13}, extra functions are incorporated to ensure the smoothness of $d(t)$ at boundary points, leading to:
\begin{equation}
\begin{aligned}
& d^{-}= \begin{cases}\dot{q}^{-}, & \text {if } q(t) \in\left[\xi_1, q^{+}\right] \\
\dot{q}^{-}\left(1-\varrho_1(q(t))\right), & \text {if } q(t) \in\left[q^{-}, \xi_1\right]\end{cases} \\
& d^{+}= \begin{cases}\dot{q}^{+}, & \text {if } q(t) \in\left[q^{-}, \xi_2\right] \\
\dot{q}^{+}\left(1-\varrho_2(q(t))\right), & \text {if } q(t) \in\left[\xi_2, q^{+}\right]\end{cases}
\end{aligned}
\label{eq27}
\end{equation}
where $\varrho_1(x)=(\sin (0.5 \pi(\sin (0.5 \pi(x-\xi_1) / \xi_3))^2))^2$, $\varrho_2(x)=(\sin (0.5 \pi(\sin (0.5 \pi(x-\xi_2) / \xi_4))^2))^2$, $\xi_1=\kappa_1 q^{-}$, $\xi_2=\kappa_2 q^{+}$, $\xi_3=q^{-}-\xi_1$, and $\xi_4=q^{+}-\xi_2$, with $0<\kappa_1, \kappa_2<1$ being two positive constants.

\begin{figure}[t]
	\centering
		\includegraphics[scale=.32]{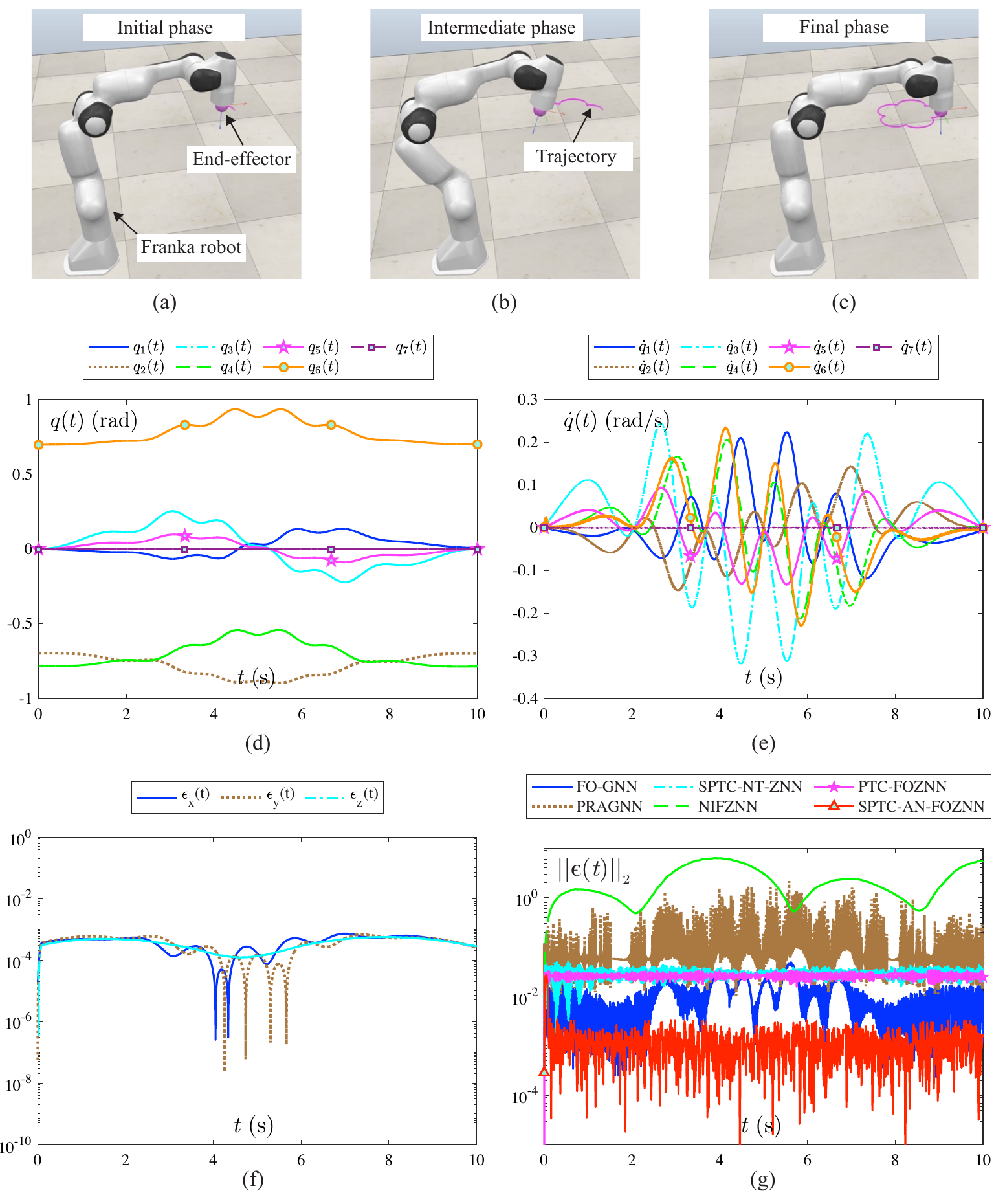}
	\caption{The snapshots of the (a) initial, (b) intermediate, and (c) final phases for a simulated Franka Emika Panda robot during tracking a five-petal-plum-shaped path, with the rendered (d) joint angles, (e) joint velocities, (f) absolute tracking errors by the SPTC-AN-FOZNN model, and (g) residual errors rendered by six different neural models with a bounded random noise $\delta(t)=\cos(t)+\bar{n}(t)$.}
	\label{fig5}
\end{figure}

\begin{figure}[t]
	\centering
		\includegraphics[scale=.32]{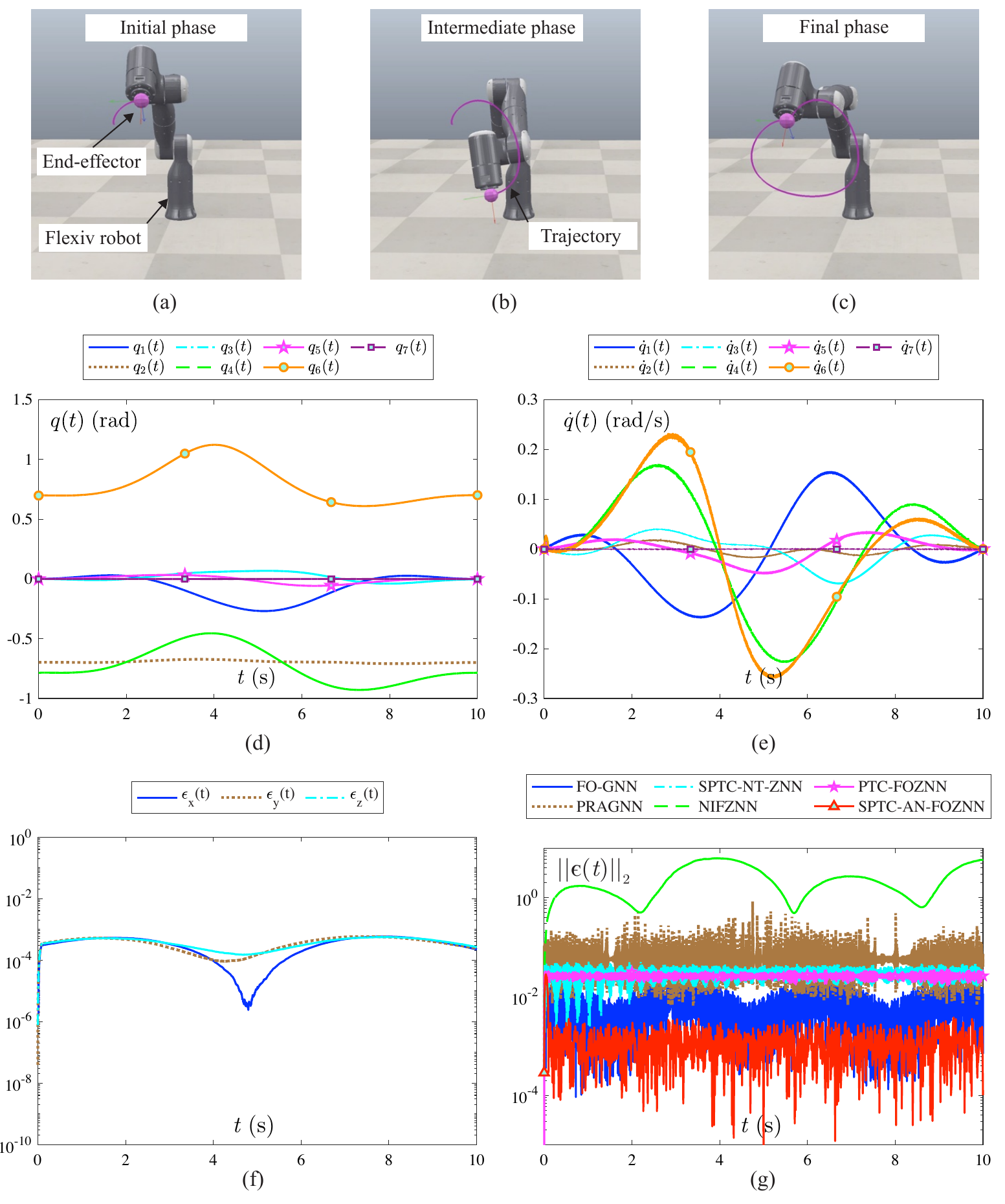}
	\caption{The snapshots of the (a) initial, (b) intermediate, and (c) final phases for a simulated Flexiv Rizon robot during tracking a heart-shaped path, with the rendered (d) joint angles, (e) joint velocities, (f) absolute tracking errors by the SPTC-AN-FOZNN model, and (g) residual errors rendered by six different neural models with a bounded random noise $\delta(t)=\cos(t)+\bar{n}(t)$.}
	\label{fig6}
\end{figure}

\begin{figure}[t]
	\centering
		\includegraphics[scale=.32]{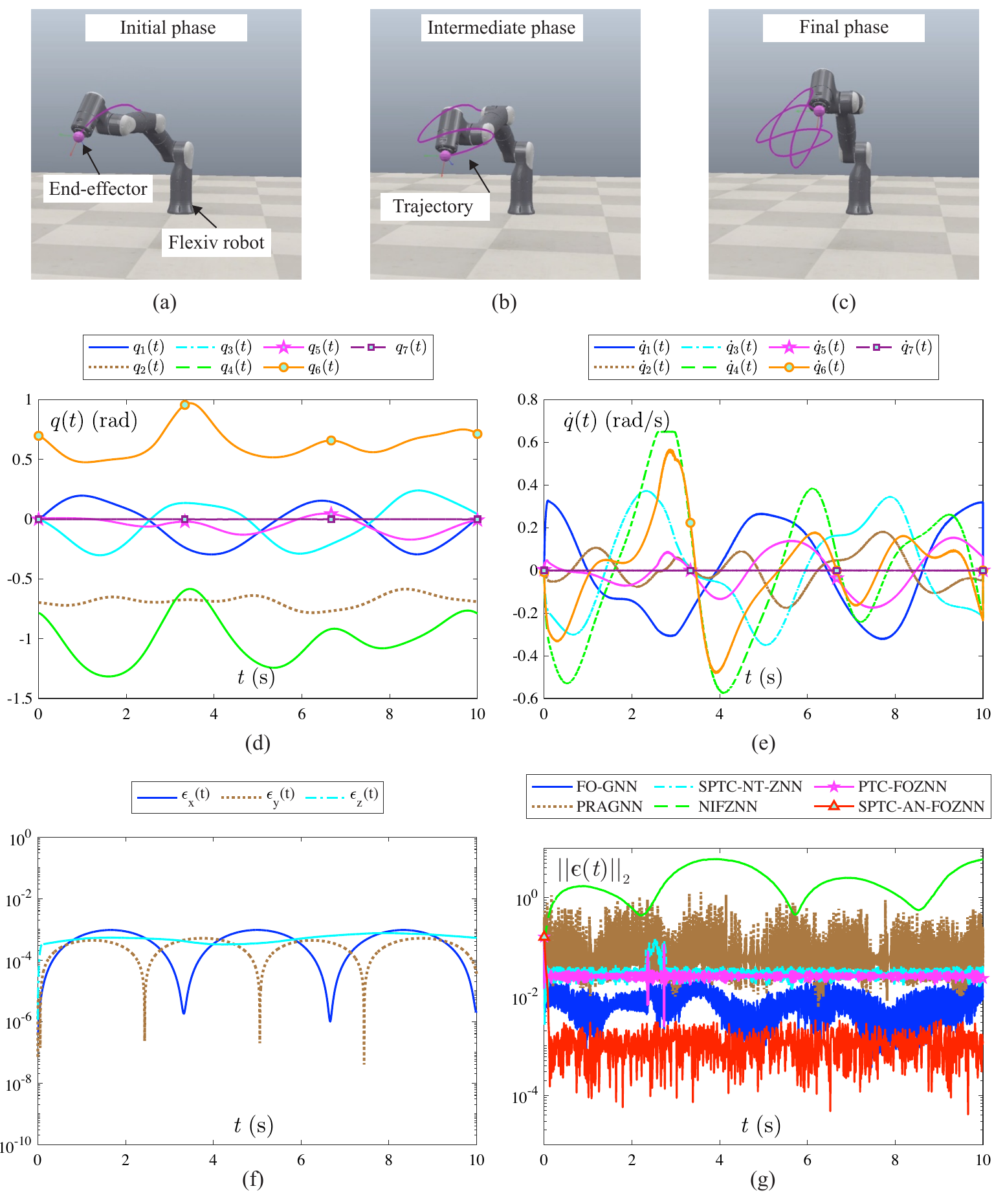}
	\caption{The snapshots of the (a) initial, (b) intermediate, and (c) final phases for a simulated Flexiv Rizon robot during tracking a Lissajous curve, with the rendered (d) joint angles, (e) joint velocities, (f) absolute tracking errors by the SPTC-AN-FOZNN model, and (g) residual errors rendered by six different neural models with a bounded random noise $\delta(t)=\cos(t)+\bar{n}(t)$.}
	\label{fig7}
\end{figure}

\begin{figure}[t]
	\centering
		\includegraphics[scale=.32]{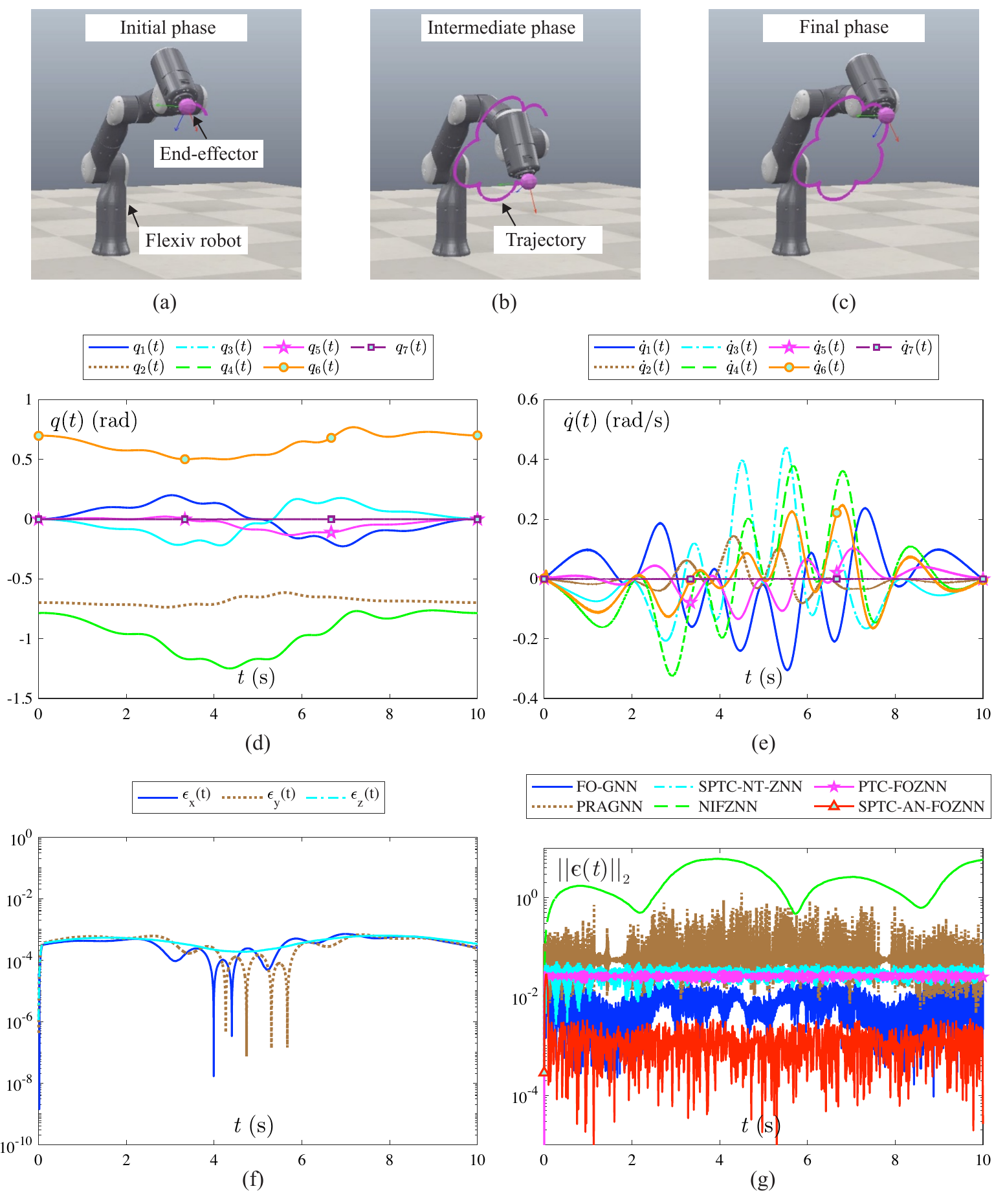}
	\caption{The snapshots of the (a) initial, (b) intermediate, and (c) final phases for a simulated Flexiv Rizon robot during tracking a five-petal-plum-shaped path, with the rendered (d) joint angles, (e) joint velocities, (f) absolute tracking errors by the SPTC-AN-FOZNN model, and (g) residual errors rendered by six different neural models with a bounded random noise $\delta(t)=\cos(t)+\bar{n}(t)$.}
	\label{fig8}
\end{figure}

\subsection{Simulation validation}
To enhance the convincibility of the simulation validation, two different types of robotic manipulators, namely the Franka Emika Panda robot (\citealp{franka2024}) and the Flexiv Rizon robot (\citealp{Murtaza2022}), are utilized to execute the tracking of three trajectory types:
\begin{itemize}
    \item Heart-shaped curve:
    \begin{equation}
\left\{\begin{array}{l}
x(t)=a(2 \sin (\theta)-\sin (2 \theta)), \\
y(t)=a(2 \cos (\theta)-\cos (2 \theta))-a, \\
z(t)=z_0,
\end{array}\right.
\label{eq28}
\end{equation}
where $a$ is a scaling factor, $z_0$ is a constant, and $\theta = 2\pi t/T$ with $T$ being the total simulation time. 
    \item Lissajous curve:
    \begin{equation}
\left\{\begin{array}{l}
x(t)=A \sin (2\pi a t/T+\delta), \\
y(t)=B \sin(2\pi bt/T), \\
z(t)=z_0,
\end{array}\right.
\label{eq29}
\end{equation}
where $A=0.06$ and $B=0.06$ are amplitudes, $a=3$ and $b=2$ are frequency factors, and $\delta=\pi/2$ is the phase difference. 
    \item Five-petal-plum-shaped curve:
    \begin{equation}
\left\{\begin{array}{l}
x(t)=r\left(\cos (\phi)+\frac{\cos (n \phi)}{n}\right), \\
y(t)=r\left(\sin (\phi)+\frac{\sin (n \phi)}{n}\right), \\
z(t)=z_0,
\end{array}\right.
\label{eq30}
\end{equation}
where $\phi=2 \pi \sin \left({\pi t}/2/T\right) \sin \left({\pi t}/2/T\right)$, $r=0.1$ is the radius, and $n=5$ is the number of lobes.
\end{itemize}

These trajectories provide a comprehensive test bed to evaluate the performance of the robotic arms under various dynamic conditions. The choice of complex trajectories like heart-shaped, Lissajous, and five-petal plum-shaped paths allows for rigorous assessment of the control algorithms in handling intricate movement patterns, crucial for real-world robotic applications.

The simulations are operationalized within the the CoppeliaSim's virtual environment (\citealp{ref32}), tackling a kinematic motion planning task with the noise-perturbed neural network controller. Specifically, the SPTC-AN-FOZNN model, configured with $\alpha=0.5$ and a predefined convergence time of $t_c=0.01$ s, addresses the control challenges posed by a random noise $\delta(t)=\cos(t)+\bar{n}(t)$, where $\bar{n}(t)$ represents the white noise bounded by 1. 

In the context of motion planning, the robot's end-effector position is specified with precision, while the other degrees of freedom remain unconstrained. The range of joint angles for the two types of robotic manipulators is set as follows:
\begin{itemize} 
    \item Lower limit ($q^-$):
    $$-[161^\circ, 131.5^\circ, 172.5^\circ, 107^\circ, 172.5^\circ, 82.5^\circ, 172.5^\circ]^\mathrm{T}$$
    \item Upper limit ($q^+$):
    $$+[161^\circ, 131.5^\circ, 172.5^\circ, 155^\circ, 172.5^\circ, 262.5^\circ, 172.5^\circ]^\mathrm{T}$$
\end{itemize}
These limits are consistent with the specified joint angle range for the real robot arm. Additionally, the joint-velocity bounds are established at $\dot{q}^-=-0.65$ rad/s and $\dot{q}^+=0.65$ rad/s, with a scaling factor $\iota$ set to $1$. This setup ensures that the robot operates within safe and efficient kinematic parameters.

The subplots (a)-(c) in Fig. \ref{fig3}-\ref{fig8} highlight the simulation results, illustrating the end-effector's precision in tracking the designated path. The subplots (d) and (e) detail the joint angles and velocities, confirming their cycling back to initial positions, thereby substantiating the SPTC-AN-FOZNN model's capability in precise kinematic motion planning. This model effectively maintains all joint movements within specified limits, with tracking errors in three dimensions kept around $10^{-4}$ meters. Additionally, the subplot (g) in Fig. \ref{fig3}-\ref{fig8} showcases the residual errors $\epsilon(t)=f(y(t),t)$ across six different RNN models, verifying the SPTC-AN-FOZNN's compliance with predefined-time convergence as theorized in Section 3.2. The comparative analysis underlines its superior tracking precision and robustness against additive noises, reinforcing its suitability for complex robotic kinematic control applications.

\subsection{Experimental validation}
\begin{figure}[t]
	\centering
		\includegraphics[scale=.285]{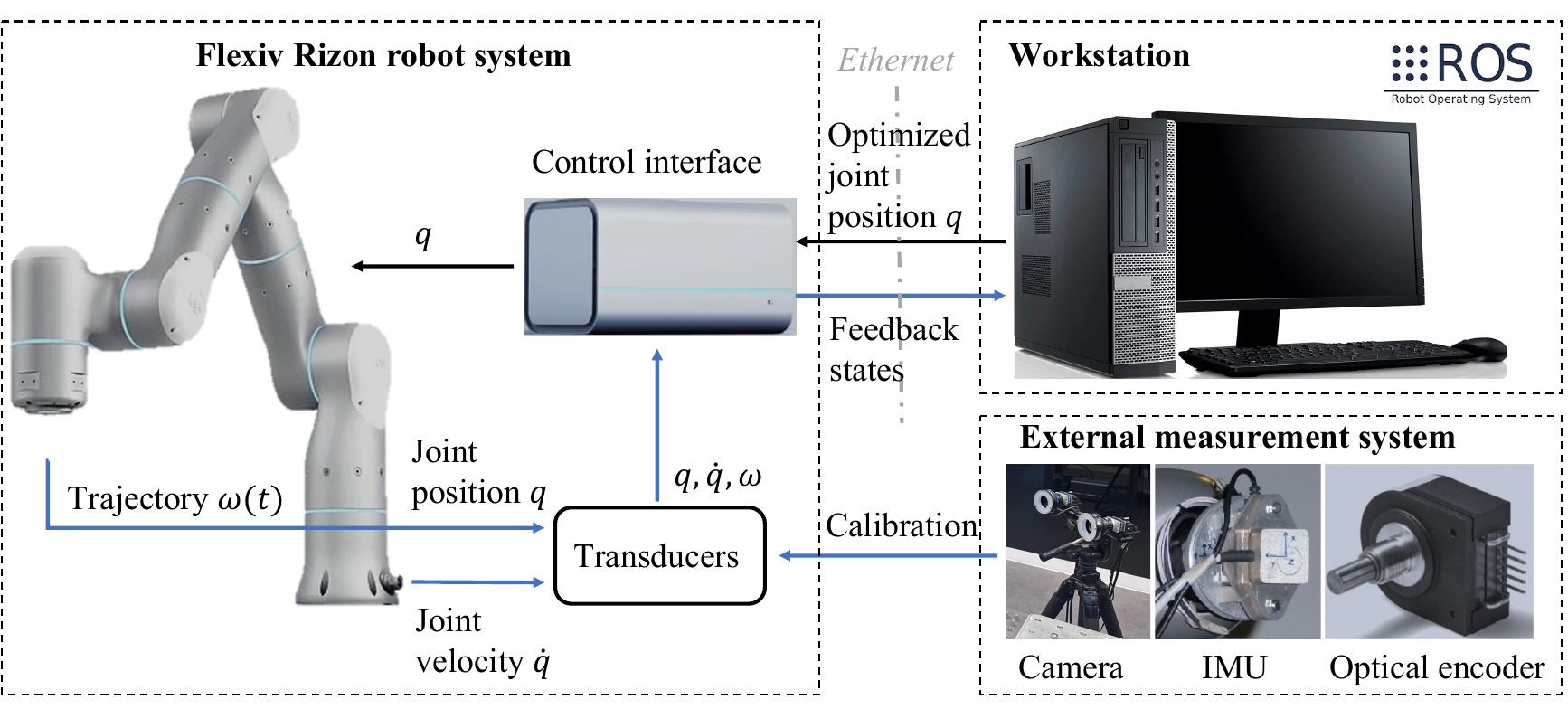}
	\caption{Schematic diagram for the experimental design of the Flexiv Rizon robot’s kinematic control system.}
	\label{fig9_pre}
\end{figure}

\textcolor{blue}{To evaluate the performance of the SPTC-AN-FOZNN model, the experimental setup incorporates the Flexiv Rizon robot system, as illustrated in the schematic diagram of Fig. \ref{fig9_pre}. The robot is controlled via a workstation running the Robot Operating System (ROS), which communicates with the control interface through an Ethernet connection. Joint positions ($q$) and velocities ($\dot{q}$) are measured using transducers, including optical encoders for angular positions of each joint, an inertial measurement unit (IMU) for dynamic joint angular velocity feedback, and an external calibrated camera for additional measurement of the end-effector’s tracking position. These sensor readings are treated as the real joint angles rather than the joint angles obtained through resolution methods. The experimental configuration considers environmental factors such as the variability in sensor readings and external disturbances. The noise model employed in the tests consists of random white noise with a sufficiently large upper bound to simulate real-world operational uncertainties. These configurations ensure the robustness of the model in scenarios involving unpredictable system dynamics and environmental disturbances.}

\begin{figure}[t]
	\centering
		\includegraphics[scale=.32]{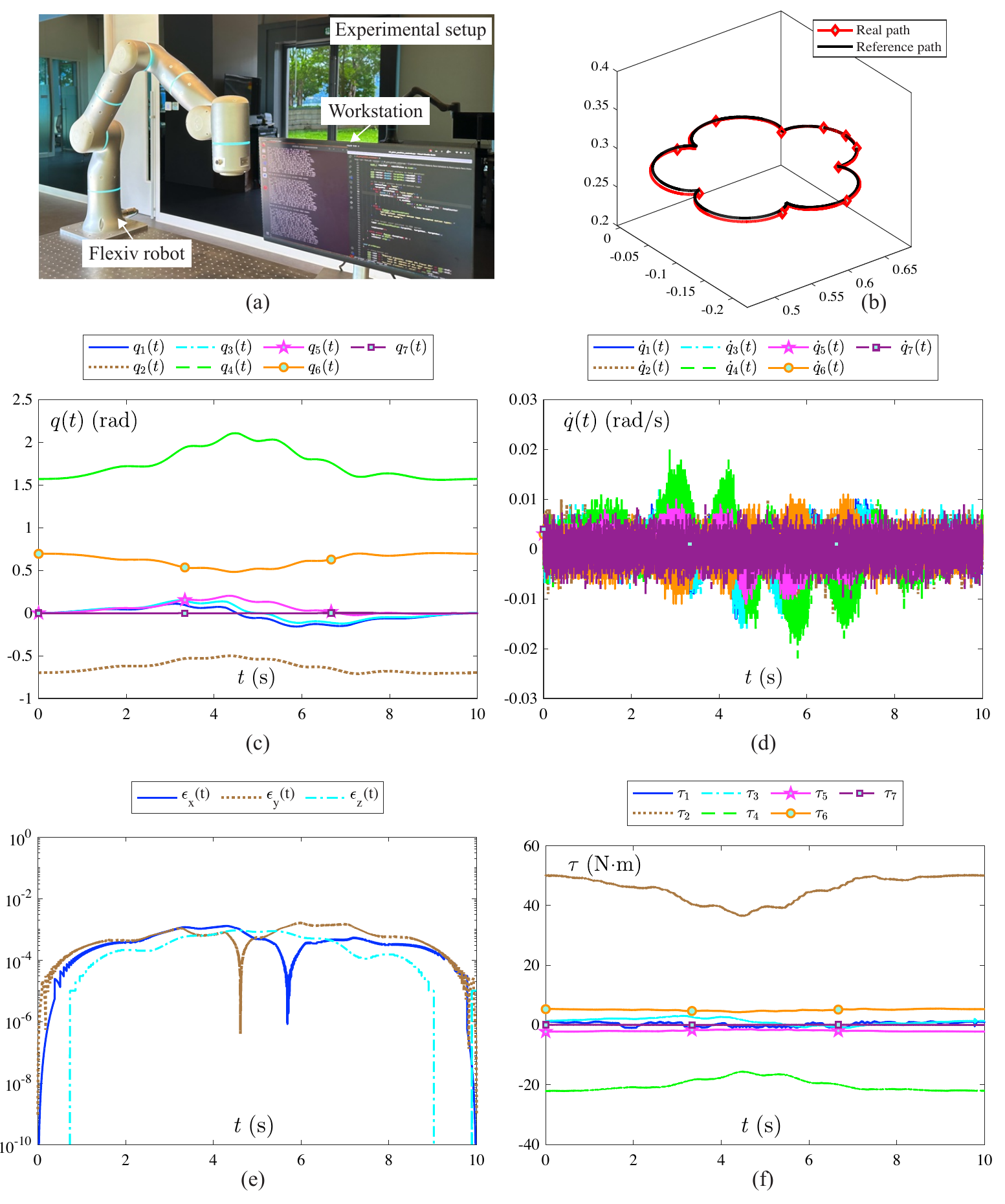}
	\caption{(a) Image of the experimental configuration, (b) comparison of the actual trajectory with the designed five-petal plum-shaped path, and the profiles of (c) joint angles, (d) joint velocities, (e) absolute tracking errors and (f) real-time joint torques produced by the SPTC-AN-FOZNN model on the actual experimental platform.}
	\label{fig9}
\end{figure}

The Flexiv Rizon robot is programmed to perform cyclical motions along a five-petal-plum-shaped trajectory, as demonstrated in Fig. \ref{fig9}(a). The measured joint angles and velocities, depicted in Fig. \ref{fig9}(c) and (d), confirm the cyclical nature of the motions, with the robot consistently returning to its initial states. Fig. \ref{fig9}(b) portrays the three-dimensional trajectory of the robot’s end-effector relative to the reference path, highlighting the achieved path-tracking accuracy ranging from $10^{-4}$ and $10^{-3}$ meters, thus aligning with the order of positional precision ($\sim0.5$ mm) calibrated in our lab and claimed in the Flexiv Rizon robot’s datasheet. In addition, the joint torques measured during the experiment are presented in Fig. \ref{fig9}(f), indicating stable and smooth motion of the robot, crucial for successful path-tracking. These results collectively affirm the SPTC-AN-FOZNN's effectiveness and its applicability in robotic motion control tasks.

\section{Conclusion}
This paper develops the SPTC-AN-FOZNN model, tailored for resolving TVQP problems, with a particular emphasis on its deployment in kinematic motion control of robots. Theoretical analysis confirmed that the SPTC-AN-FOZNN model achieves strictly predefined-time convergence and exhibits anti-noise characteristics. Numerical assessments involving an illustrative TVQP example across six distinct RNN models, demonstrated that the SPTC-AN-FOZNN model outperforms other RNNs in terms of convergence precision and robustness. The practicability of the SPTC-AN-FOZNN model is further underscored by its successful application to the kinematic motion control of a Franka Emika Panda robot and a Flexiv Rizon robot. This marks a seminal demonstration of a strictly predefined-time convergent ZNN model with time-diminishing variable gain. This innovation suggests promising future directions for more energy-efficient ZNN hardware architectures exhibiting strictly user-prescribed-time convergence.

\textcolor{blue}{While effective, the current model has limitations that require further exploration. Its reliance on fixed parameters, like $\alpha$ values and noise properties, may restrict its use in varied environments or complex systems. It also assumes predictable and uniform noise, which is often unrealistic in dynamic, real-world scenarios. Future work could focus on adaptive parameter tuning to improve the model's responsiveness to abrupt changes in system dynamics or noise conditions.}

\section*{Funding} 
This work was supported by the National Natural Science Foundation of China under Grant 52205032, in part by the Shun Hing Institute of Advanced Engineering, The Chinese University of Hong Kong, and in part by Research Grants Council of Hong Kong (Ref. No. 14204423).
\section*{Ethical approval}
Not required.
\section*{CRediT authorship contribution statement}
\textbf{xx:} Conceptualization, Methodology, Investigation, Writing – review \& editing. \textbf{xx:} Software, Investigation, Writing– original draft. \textbf{xx:} Visualization, Supervision, Funding
acquisition. \textbf{xx:} review. \textbf{xx:} Investigation. \textbf{xx:} Investigation, Supervision.

\section*{Declaration of competing interest} The authors declare that they have no known competing financial interests or personal relationships that could have appeared to
influence the work reported in this paper



\bibliographystyle{apalike2}

\bibliography{cas-refs}

\end{document}